\documentclass[twocolumn,groupedaddress,aps,pra,
amsmath,amssymb,showpacs]{revtex4-1}
\usepackage[colorlinks,linkcolor=blue,anchorcolor=blue,
            citecolor=blue,bookmarksnumbered]{hyperref}
\usepackage{graphicx,epsfig,float,epstopdf}
\usepackage{txfonts}
\usepackage{amssymb,amsmath,subeqnarray,mathrsfs,
           amsfonts,bm,makecell,cases,color,extpfeil,lipsum}
\usepackage{dcolumn,comment}\allowdisplaybreaks

\begin{document}
\title{Quantum Squeezing of Slow-Light Dark Solitons via Electromagnetically Induced Transparency}
\author{Jinzhong Zhu$^1$ and Guoxiang Huang$^{1,2,3}$}
\affiliation{$^1$State Key Laboratory of Precision Spectroscopy,
                 East China Normal University, Shanghai 200241, China\\
             $^2$NYU-ECNU Joint Institute of Physics, New York University  at Shanghai, Shanghai 200062, China\\
             $^3$Collaborative Innovation Center of Extreme Optics, Shangri University, Taiyuan 030006, China
             }

\begin{abstract}

We consider the quantum effect of slow light dark soliton (SLDS) in a cold atomic gas with defocuing Kerr nonlinearity via electromagnetically induced transparency (EIT). We calculate the quantum fluctuations of the SLDS by solving the relevant non-Hermitian eigenvalue problem describing the quantum fluctuations, and find that only one zero mode is allowed. This is different from the quantum fluctuations of bright solitons, where two independent zero modes occur. We rigorously prove that the eigenmodes, which consist of continuous modes and the zero mode, are bi-orthogonal and constitute a complete bi-orthonormalized basis, useful for the calculation on the quantum fluctuations of the SLDS. We demonstrate that, due to the large Kerr nonlinearity contributed from the EIT effect, a significant quantum squeezing of the SLDS can be realized; the squeezing efficiency can be manipulated by the Kerr nonlinearity and the soliton's amplitude, which can be much higher than that of bright solitons. Our work contributes to efforts for developing quantum nonlinear optics and non-Hermitian Physics, and for possible applications in quantum information processing and precision measurements.

\end{abstract}

\maketitle

\section{Introduction}

Optical dark pulses, localized dips (or holes) on a homogeneous bright background, have received much attention in classical and quantum optics. Comparing with bright pulses, they possess many attractive advantages, including being more stable and less sensitive to noise~\cite{Agrawal2019,Kivshar2003}, which are desirable for information processing and transformation and hence play significant roles in many research fields of physics~\cite{krokel,Gredeskul,Meshulach,Cao,Schoenlein,Farina,Xue,Bouldja}. One of typical example of dark pulses is optical dark soliton, formed by the balance between dispersion and defocusing Kerr nonlinearity~\cite{Agrawal2019,Kivshar2003}.

In recent years, many efforts have been paid to the research on electromagnetically induced transparency (EIT), an important quantum interference effect typically occurring in a $\Lambda$-type three-level atomic system that interacts resonantly with two laser fields. EIT can be used to suppress resonant optical absorption, slow down group velocity, enhance Kerr nonlinearity, etc., and hence has tremendous practical applications~\cite{Fleischhauer2005,Khurgin2009}. Interestingly, EIT systems support slow-light solitons~\cite{Deng2004PRL,Huang2005PRE,Deng2004OL,Hang2006,Michinel2006,Qi2011,Khadka2014,Facao2015,
Bai2019}, which can be
manipulated actively~\cite{Bai2019,Bai2013,Chen2014}. However, up to now
most works were mainly focused on slow-light bright solitons, and limited in semi-classical regime~\cite{Shou2020}.

In this article, we investigate the quantum effect of optical dark pulses in a cold, three-level atomic gas working on the condition of EIT. By using suitable one- and two-photon detunings, a large defocuing Kerr nonlinearity and hence slow-light dark solitons (SLDS) can be generated in the system. The expression of the quantum fluctuations of the SLDS is obtained through solving Bogoliubov-de Gennes (BdG) equations, which are non-Hermitian eigenvalue problem describing the quantum fluctuations. We find that this eigenvalue problem allows only a single zero mode (i.e. eigenmode with zero eigenvalue), which is different from the case of bright-soliton fluctuations where there are two independent zero modes.

Based on the above results, we rigorously prove that the all eigenmodes, consisting of continuous (Goldstone) modes and the zero mode, are bi-orthogonal and constitute a complete bi-orthonormal basis, by which the quantum dynamics of the SLDS is studied analytically.
We demonstrate that a significant quantum squeezing of the SLDS can be realized, which is originated from the large Kerr nonlinearity of the system. Moreover, the squeezing efficiency can be manipulated by the Kerr nonlinearity and the soliton amplitude; interestingly, the squeezing efficiency of the SLDS is higher than that of bright solitons. The method and results presented here are useful for developing quantum nonlinear optics and non-Hermitian physics~\cite{Vladimir2016,El-Ganainy2017,Ashida2020,Bergholtz2021}, and may be applied to the study of Bose-Einstein condensation, quantum information processing, and precision measurements, etc.

Before preceding, we stress that, although a large amount of studies on quantum optical solitons were reported in the past years, our work is different from them, with the reasons given in the following:

(i)~Most studies on quantum effects of optical solitons reported so far were devoted to the bright solitons in optical fibers~\cite{DrummondPRL1987,Potasek1988,HausPRA1989-1,HausPRA1989-2,HausJOSAB1990,
Rosenbluh1991,Drummond1993,YLai1993,YLai1995,Duan1995,YaoPRA2,Hagelstein1996,Margalit1998,
Yeang1999,Haus2000,Matsko2000,Drummond2001,Corney2001,Kumar2002,Kozlov2002,
RKLee2004,Rand2005-1,Rand2005-2,Corney2006,Tsang2006,Lai2009,Tran2011,Honarasa2011,
Drummond2014,Hosaka2016,Zou2021}. The main analytical approach used is soliton perturbation theory.
Generally, perturbations on solitons have contributions from both
zero modes  and continuous modes, but most of these studies considered zero modes only. Although in Refs.~\cite{Margalit1998, Haus2000} continuous modes have been taken into account, the completeness and orthonormality of the eigenmode set were not discussed~\cite{note1}. In our work, the all eigenmodes are obtained, and their completeness and orthonormality are proved rigorously for the first time.

(ii)~The quantum effect of dark solitons in optical fibers was investigated in Refs.~\cite{Corney2001,Honarasa2011}; however, the contribution of continuous modes was not considered there. In addition, the zero modes, as did for bright solitons~\cite{HausJOSAB1990,YLai1993,YLai1995,Yeang1999,Haus2000,Corney2001,
Rand2005-1,Rand2005-2}, was determined by a phenomenological method (i.e. they are obtained by simply taking the derivatives of the soliton solution with respect to the free parameters in the solution). Such a method and related results obtained are uncomplete or incorrect because the zero modes obtained are generally not independent (see the discussion in Sec.~\ref{section3b} below). Differently, in our work the eigenmodes are acquired by solving the eigenvalue problem of the perturbation, and the zero modes are determined by the requirement of the completeness and orthonormality of the all eigenmodes. The results presented in our work provides a clear way to avoid puzzles and confusions on zero mode problem in previous studies~\cite{Corney2001,Bilas2005,Yu2007,Honarasa2011,Sykes2011,Walczak2012,
Takahashi2015} of perturbation and quantum effects of dark solitons.

(iii)~At variance with Refs.~\cite{DrummondPRL1987,Potasek1988,HausPRA1989-1,HausPRA1989-2,HausJOSAB1990,
Rosenbluh1991,Drummond1993,YLai1993,YLai1995,Duan1995,YaoPRA2,Hagelstein1996,Margalit1998,
Yeang1999,Haus2000,Matsko2000,Drummond2001,Corney2001,Kumar2002,Kozlov2002,
RKLee2004,Rand2005-1,Rand2005-2,Corney2006,Tsang2006,Lai2009,Tran2011,Honarasa2011,
Drummond2014,Hosaka2016,Zou2021}, which are for the quantum solitons in optical fibers, the work reported here is on the quantum effect of SLDS generated in a cold atomic gas via EIT. Our work can be taken as an extension of a recent publication~\cite{Zhu2021}, but it is not a simple extension because the perturbation approach on the SLDS is quite different from that of slow-light bright solitons. One of the main differences is that the SLDS has a  continuous background, which makes the eigenvalue problem of the perturbation be very different from that of the slow-light bright solitons. In particular, the system allows only a single zero mode (not like the case of slow-light bright solitons where two independent zero modes occur) and  the quantum squeezing of the SLDS displays behavior different from that of the slow-light bright solitons.

The reminder of the article is organized as follows. In Sec.~\ref{Sec2}, we present the model under study and the quantum nonlinear Schr\"odinger (QNLS) equation describing the nonlinear evolution of quantized probe laser field.  In Sec.~\ref{Sec3}, we diagonalize the effective Hamiltonian by expressing the quantum fluctuations of the SLDS as a superposition of the complete and bi-orthonormalized eigenmodes, obtained by solving the eigenvalue problem (BdG equations) of the quantum fluctuations. In Sec.~\ref{Sec4}, the quantum squeezing of the SLDS is investigated in detail. Lastly, Sec.~\ref{Sec5} gives a summary of the main results obtained in this work.

\section{Model and envelope equation for the propagation of the quantized probe field }\label{Sec2}

\subsection{Physical model}\label{section2a}

\begin{figure*}
\includegraphics[width=0.92\textwidth]{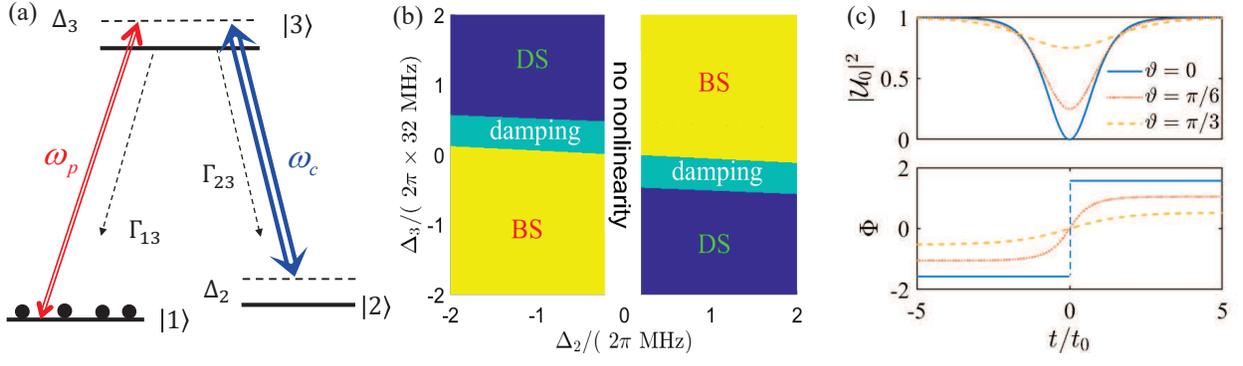}
\caption{(a)~Excitation scheme of the EIT-based $\Lambda$-type atomic gas. Solid black dots mean that the atoms are initially prepared at the ground state $|1\rangle$.  For more detail, see text.
(b)~Existence regions of slow-light dark and bright solitons in
the plane of one-photon detuning $\Delta_3$ and two-photon detuning $\Delta_2$.
Yellow: the region where bright solitons exist;
purple: the region where dark solitons exist;
cyan: the region where the damping is dominant over the dispersion and Kerr nonlinearity;
white: the region where Kerr nonlinearity plays no significant role.
In both the cyan and white regions, no solitons exist.
(c)~Dark soliton intensity (upper) $|{\cal U}_{0}|^2$ and phase (lower) $\Phi$
as functions of dimensionless time $t/t_0$ for
different blackness parameter $\vartheta$.
}
\label{Fig1}
\end{figure*}

The system we consider is a cold atomic gas interacting with a probe and a control laser fields, forming a standard $\Lambda$-type three-level configuration [see Fig.~~\ref{Fig1}(a)].
In the system, $|1\rangle$ and $|2\rangle$ are
two nearly degenerate ground states, and $|3\rangle$ is an excited state with spontaneous-emission decay rates $\Gamma_{\alpha3}(\alpha=1,2)$ to $|1\rangle$ and $|2\rangle$, respectively. The probe field  is weak and pulsed (with center angular frequency $\omega_p$ and wavenumber $k_p=\omega_p/c$; $c$ is light speed in vacuum), coupling to the transition $|1\rangle\leftrightarrow|3\rangle$; control laser field  is a strong continuous-wave (with angular frequency $\omega_c$  and wavenumber $k_c=\omega_c/c$), coupling to the transition $|2\rangle\leftrightarrow|3\rangle$.
$\Delta_{2}$ and $\Delta_{3}$ are two- and one-photon detunings, respectively. For suppressing Doppler effect,
both the probe and control fields are assumed to propagate along the same (i.e. $z$) direction.

For simplicity, we assume the atomic gas is cigar-shaped, which can be realized by filling it  into a waveguide or by taking the transverse distribution of the probe field to be large enough so that the diffraction effect can be neglected. Thus one can use a reduced (1+1)-dimensional model to describe the probe-filed propagation, with the total electric field given by
${\hat{\bf E}}(z,t)={\bf E}_{c}(z,t)+{\hat{\bf E}}_{p}(z,t)$. Here,
${\bf E}_{c}(z,t)\equiv{\bf e}_{c}{\cal E}_{c}(z,t)e^{i(k_{c}z-\omega_{c}t)}+{\rm c.c.}$ and
${\hat{\bf E}}_{p}(z,t)\equiv{\bf e}_{p}{\cal E}_{p}\hat{E}_{p}(z,t)e^{i(k_{p}z-\omega_{p}t)}+{\rm H.c.}$  are respectively the quantized probe and c-number control fields, with c.c. (H.c.) representing the conjugate (Hermitian conjugate); ${\bf e}_{c}$ and ${\cal E}_{c}$  are respectively the unit polarization vector and the amplitude of the control field;
${\bf e}_{p}$ and ${\cal E}_{p}\equiv\sqrt{\hbar\omega_{p}/(2\varepsilon_{0}V)}$\, ($V$ is quantized volume) are respectively unit polarization vector and the single-photon amplitude of the probe field.  $\hat{E}_{p}(z,t)$
is a slowly-varying annihilation operator of probe photons, obeying the commutation relation $[\hat{E}_{p}(z,t),\hat{E}_{p}^{\dag}(z',t)]=L\delta(z'-z)$, with $L$ the quantization length along
the $z$-axis.

Under electric-dipole and paraxial approximations, the system Hamiltonian is given by
${\hat H}=\int dz\,\,\bigg[-\frac{\hbar c}{L}\hat{E}_{p}^{\dag}\left(i\frac{\partial}{\partial z}\right)\hat{E}_{p} -\frac{\hbar N}{L}\left(\sum_{\alpha=2,3}\Delta_{\alpha}\hat{S}_{\alpha\alpha}\right.
\left.+g_{p}\hat{S}_{31}^\dag\hat{E}_{p}+\Omega_c\hat{S}_{32}^\dag+{\rm H.c.}\right)\bigg]$.
Here $N$ is the total atomic number of the system; $\hat{S}_{\alpha\beta}(z,t)=
\hat{\sigma}_{\beta\alpha}\,e^{i[(k_\beta-k_\alpha)z-(\omega_\beta-\omega_\alpha+\Delta_\beta-\Delta_\alpha)t]}$ $(\alpha,\beta = 1,2,3)$
are atomic transition operators;
$\Omega_c=(\mathbf{e}_c\cdot \mathbf{p}_{32})\mathcal{E}_c/\hbar$ is the half Rabi frequency of the control field; $g_{p}=(\mathbf{e}_p\cdot\mathbf{p}_{31})\mathcal{E}_p/\hbar$ is single-photon half Rabi frequency of the probe field;  $\mathbf{p}_{\alpha\beta}$ is the electric dipole matrix element associated with the transition from $|\beta\rangle$ to $|\alpha\rangle$; the detunings are defined by $\Delta_{2}=\omega_{p}-\omega_{c}-(\omega_{2}-\omega_{1})$,  $\Delta_{3}=\omega_{p}-(\omega_{3}-\omega_{1})$.

The dynamics of the system is governed by the Heisenberg-Langevin and the Maxwell (HLM) equations, given by
\begin{subequations}\label{HLM}
\begin{align}
& \frac{\partial}{\partial t}{\hat S}_{\alpha\beta}=\frac{1}{i\hbar}\left[{\hat S}_{\alpha\beta},{\hat H}\right]-{\hat{\cal L}}({\hat S}_{\alpha\beta})+{\hat F}_{\alpha\beta}, \label{HLM1}\\
& i\left(\frac{\partial}{\partial z}+\frac{1}{c}\frac{\partial}{\partial t}\right)\hat{E}_{p}+\frac{g_{p}^\ast N}{c}{\hat S}_{31}=0, \label{HLM2}
\end{align}
\end{subequations}
where ${\hat{\cal L}}\,({\hat{S}_{\alpha\beta}})$ is the $3\times3$ relaxation matrix including the atomic decay rates of the spontaneous emission and dephasing, ${\hat F}_{\alpha\beta}$ are $\delta$-correlated Langevin noise operators introduced to preserve the Heisenberg commutation relations for the operators of the atoms and the probe field. Explicit expressions of Eq.~(\ref{HLM}a) are presented in Appendix~\ref{app1}.

The model described above can be realized by many atomic systems. One of the candidates is the laser-cooled alkali $^{87}$Rb gas, with the levels chosen to be $|1\rangle=|5^2S_{1/2},F=1,m_{F}=1\rangle$, $|2\rangle=|5^2S_{1/2},F=2,m_{F}=1\rangle$ and $|3\rangle=|5^2P_{3/2},F=2,m_{F}=1\rangle$, with $\Gamma_{13}=\Gamma_{23}\approx 2\pi\times3\,{\rm MHz}$~\cite{Steck}, which will be used below.

\subsection{Nonlinear envelope equation and the existence region of dark solitons in parameter space}\label{section2b}

To understand the quantum dynamics of the system, we need to solve the nonlinearly coupled equations (\ref{HLM1}) and~(\ref{HLM2}), which have many degrees of freedom of atoms and photons and thus not easy to approach. A convenient way is to reduce such equations to an effective one by eliminating the atomic degrees of freedom  under some approximations, which has been recently used to the study of polaritons in atomic gases~\cite{Larre2015,Larre2016,Gullans2016}.
Similar to Ref.~\cite{Zhu2021}, by employing the perturbation expansion under weak-dispersion and weak-nonlinearity approximations, one can obtain the following QNLS equation
\begin{align}\label{QNLS0}
& i\left[\left(\frac{\partial}{\partial z}+\frac{1}{V_{g}}\frac{\partial}{\partial t}\right)+{\rm Im}(K_{0})\right]\hat{E}_{p}-\frac{K_{2}}{2}\frac{\partial^2}{\partial t^2}\hat{E}_{p} \notag\\
&\hspace{2 cm} +W|g_p|^{2}\hat{E}_{p}^{\dag}\hat{E}_{p}\hat{E}_{p}-i{\hat{\cal F}}_{p}e^{-i\tilde{K_0}z}=0,
\end{align}
which describes the nonlinear evolution of the probe-field envelope $\hat{E}_{p}$. Here, $K_j\equiv
(\partial^j K/\partial \omega^j)^{-1}|_{\omega=0}$ ($j=0,1,2$), with $K=K(\omega)$ the linear dispersion
relation, $V_g\equiv 1/K_1$ the group velocity, and  $K_2$ the coefficient of the group-velocity dispersion;
$W$ is the coefficient of third-order Kerr nonlinearity, which is proportional to the third-order nonlinear optical susceptibility $\chi_{p}^{(3)}$; ${\hat{\cal F}}_{p}(z,t)$ is the $\delta$-correlated induced Langevin noise operator. For explicit expressions of $K(\omega)$,  $W$, and ${\hat{\cal F}}_{p}(z,t)$, see Appendix~\ref{app2}.
{\color{blue}Note that, in general, the coefficients in Eq.~(\ref{QNLS0}) depend on $\omega$ (i.e. the sideband frequency of the probe pulse). Because we are interested in the propagation of probe pulse with the center frequency $\omega_p$, the coefficients in Eq.~(\ref{QNLS0}) are estimated at $\omega=0$. In this situation, these coefficients ($K_2$, $W$, etc.) are functions of the one- and two-photon detunings (i.e. $\Delta_3$ and $\Delta_2$), and other system parameters (see Ref.~\cite{Zhu2021}).
}

The coefficients of Eq.~(\ref{QNLS0}) are generally complex  due to the near-resonant character of the system. However, under the condition of EIT (i.e. $|\Omega_c|^2\gg \gamma_{21}\gamma_{31}$), the imaginary part of these coefficients are much smaller than their real parts and hence can be neglected.
Depending on the sign of $W/K_2$, in the case of classical limit and negligible noise, Eq.~(\ref{QNLS0}) admits dark (bright) soliton solutions for $W/K_2>0~(W/K_2<0)$.

Fig.~\ref{Fig1}(b) shows the existence regions of dark and bright solitons in the parameter space with $\Delta_3$  and  $\Delta_2$  as two coordinates. In the figure, the purple region ``DS'' (yellow region ``BS'') is  the one where dark (bright) solitons exist. The cyan region indicated by ``damping''  is the one where the damping of the probe field is dominant over the group-velocity dispersion and Kerr nonlinearity; the white region indicated by ``no nonlinearity'' means that in this domain the Kerr nonlinearity plays no significant role. Thus, in both the cyan and white regions, the system does not support soliton~\cite{Shou2020}. When plotting the figure, we have taken $\Gamma_{13}=\Gamma_{23}\approx 2\pi\times3\,{\rm MHz}$,
${\cal N}_a$ (atomic density)$=8.8\times10^{11}\,{\rm cm}^{-3}$, $|g_{p}|^2N/c=2.4\times10^{10}\,{\rm cm^{-1}s^{-1}}$, $\Omega_{c}=2\pi\times42\,{\rm MHz}$, and $t_{0}\,\, {\rm (the\,\, time\,\, duration\,\, of\,\, the\,\, probe\,\, pulse)}=5.5\times10^{-8}\,{\rm s}$.

{\color{blue}
Since the atomic gas is nearly resonant with the probe and control fields and works under the condition of EIT, the system can possess a large Kerr nonlinearity. As an example, by taking $\Delta_{2}=-2\pi\times1.6 \,{\rm MHz}$, $\Delta_{3}=2\pi\times64\,{\rm MHz}$, 
and using the formulas for the linear dispersion relation 
$K$ and the Kerr coefficient $W$ given in the Appendix~\ref{app2},
we obtain  $K_{1}\approx 3.08\times10^{-7}\,{\rm cm^{-1}s}$,
$K_{2}\approx3.19\times10^{-15}\,{\rm cm^{-1}s^2}$, $W\approx 8.20\times10^{-17}\,{\rm cm^{-1}s^2}$. Thus we have
\begin{equation}\label{chi3}
\chi_p^{(3)}={\color{blue}\frac{2c|{\bf e}_{p}\cdot\mathbf{p}_{31}|^{2}}{\hbar^2\omega_{p}}W}\approx 1.20 \times 10^{-10}\,{\rm m}^2{\rm V}^{-2}.
\end{equation}
Because $\chi_p^{(3)}$ is proportional to $\Delta_2$, so non-zero two-photon detuning (i.e. $\Delta_2\neq 0$) is necessary to obtain the large Kerr nonlinearity.
Such a Kerr nonlinearity, which is more than ten orders of magnitude larger than that of conventional optical media (such as optical fibers)~\cite{Agrawal2019}, is the main reason why optical solitons can form at very low-light level in EIT-based atomic gases~\cite{Deng2004PRL,Huang2005PRE,Deng2004OL,Hang2006,Michinel2006,Qi2011,
Khadka2014,Facao2015,Bai2019,Bai2019,Bai2013,Chen2014,Shou2020}.

Note that even for the large Kerr nonlinearity given above, the perturbation expansion used for deriving the QNLS equation (\ref{QNLS0}) can still be applied. The reasons are the following: in our consideration the light intensity of the probe pulse is small, and its time duration is large (which means that  its dispersion is weak). Thus, the perturbation expansion is obtained under weak-dispersion and weak-nonlinearity approximations. In fact, similar perturbation expansion was used in Refs.~[13-24].
}

After neglecting the imaginary parts of $K_1$, $K_2$, and $W$, Eq.~(\ref{QNLS0}) can be written as the dimensionless form
$i\frac{\partial}{\partial s}\hat{U}+\frac{\partial^2}{\partial\tau^2}\hat{U}-2g\hat{U}^{\dag}\hat{U}\hat{U}
=-2i\nu\hat{U}+i\hat{f}_{p}$,
with $\hat{U}=\hat{E}_{p}/\sqrt{n_{0}}$  ($n_{0}\gg 1$ is typical mean photon number in the probe field~\cite{note2}), $s=z/(2L_{\rm disp})$, $\tau=(t-z/V_{g})/t_{0}$, ${\hat f}_{p}=2L_{\rm disp}{\hat{\cal F}}_{p}e^{-iK_0z}$, $\nu=L_{\rm disp}/L_{\rm abs}$, and  $g=L_{\rm disp}/L_{\rm nln}$ (the dimensionless parameter characterizes the magnitude of the Kerr nonlinearity). Here, $L_{\rm disp}\equiv t_{0}^2/|K_2|$, $L_{\rm nln}\equiv [n_{0} |g_{p}|^2|W|]^{-1}$, and $L_{\rm abs}\equiv 1/{\rm Im}(K_{0})$ are typical dispersion length, nonlinearity length, and absorption length of the probe field, respectively.

Due to the EIT effect and the ultracold environment, the Langevin noise operators make no contribution to the normally-ordered correlation functions of system operators~\cite{Gorshkov20071,Gorshkov20072}, also the dimensionless absorption coefficient $\nu\approx1.69\times10^{-2}\ll1$. Taking account these facts and making the transformation $\hat{U}=\hat{{\bar U}}e^{-i\mu s}$, we obtain the reduced QNLS equation
\begin{eqnarray}\label{QNLSE2}
i\frac{\partial}{\partial s}\hat{\bar{U}}=
 -\frac{\partial^2}{\partial\tau^2}\hat{\bar{U}}+2g\hat{\bar U}^{\dag}\hat{\bar{U}}\hat{\bar{U}}-\mu\hat{\bar{U}},
\end{eqnarray}
with the  parameter $\mu$ the ``chemical potential'' to be specified lately.
The {\it effective} Hamiltonian for the system described by the QNLS equation (\ref{QNLSE2}) reads
\begin{eqnarray}\label{EffectiveH}
& \hat{H}_{\rm eff}=\int_{-\infty}^{+\infty}d\tau{\hat{\bar U}}^{\dagger}
  \left(-\frac{\partial^2}{\partial\tau^2}-\mu-g{\hat{\bar U}}^{\dag}{\hat{\bar U}}\right){\hat{\bar U}}.
\end{eqnarray}

\section{Complete and bi-orthonormal set of the eigenmodes for the quantum fluctuations of slow-light dark solitons}\label{Sec3}

\subsection{Slow-light dark solitons}\label{section3a}

Our main aim is to investigate the quantum fluctuations from classical dark solitons. As a first step, we consider the classical limit of the system, which is valid when the probe field contains a large photon number. In this case, the operator $\hat{\bar U}$ can be approximated by a c-number function ${\cal U}_0$.  Then the reduced QNLS Eq.~(\ref{QNLSE2}) becomes a classical NLS equation of the form $i\partial  {\cal U}_0/\partial s+\partial^2 {\cal U}_0/\partial\tau^2-2g|{\cal U}_0|^2 {\cal U}_0+\mu {\cal U}_0=0$.
When working in the ``DS'' region of Fig.~\ref{Fig2}(b), this equation admits the fundamental dark-soliton solution
\begin{align}\label{CS0}
{\cal U}_0(s,\tau)&={\cal A}\sqrt{g}\left(\cos\vartheta\tanh\sigma+i\sin\vartheta\right)e^{i\theta_{0}},
\end{align}
with $\sigma={\cal A}g\cos\vartheta\left(\tau-\tau_{0}-2{\cal A}g\sin\vartheta s\right)$ and $\mu=2{\cal A}^{2}g^{2}$.
Here, ${\cal A}$ and $\theta_0$ are constants characterizing the amplitude and the overall phase of the soliton; $\vartheta$ ($0\leq\vartheta\leq\pi/2$) is a constant characterizing the dark-soliton blackness, defined by ${\cal A}^{2}g\cos^2\vartheta$ (i.e. the difference between the minimum of the soliton intensity and the background intensity ${\cal A}^{2}g$); the ``momentum'' and initial ``position'' of the soliton are given by  $2{\cal A}g\sin\vartheta$ and $\tau_0$, respectively. The soliton for the special case $\vartheta=0$ is called black soliton; in general case ($\vartheta\neq0$), it is called the dark (or grey) soliton. Fig.~\ref{Fig1}(c) shows the profile of the dark soliton intensity  $|{\cal U}_{0}|^2$ (upper part) and phase $\Phi\equiv \arctan(\cos\vartheta\tanh\sigma/\sin\vartheta)$ (lower part) with different values of the blackness parameter $\vartheta$. When plotting the figure, we have taken ${\cal A}=g=1$, and
hence the background intensity of the dark soliton is ${\cal A}^{2}g=1$.

By using the physical parameters given in Sec.~\ref{section2b}, we can estimate the propagation velocity of the dark soliton, given by
\begin{align}\label{Vsol}
V_{\rm sol}=V_g+\frac{{\cal A}g t_0\sin \vartheta}{L_{\rm disp}}
\approx 1.08\times10^{-4} c,
\end{align}
for ${\cal A}=1$, $t_0=5.5\times10^{-8} s$, and $\vartheta=\pi/2$. We see that the soliton velocity is much smaller than $c$ (i.e. it indeed is a SLDS), which is due to the EIT effect induced by the control field.

\subsection{Eigenmodes of the quantum fluctuations and their bi-orthogonality and completeness}\label{section3b}

\subsubsection{Eigenvalue problem of the quantum fluctuations}

Now we consider the quantum correction of the SLDS solution (\ref{CS0}). We assume the mean photon number $n_0$ in the probe field is much larger than one, the quantum fluctuations of the SLDS are weak, and hence we can take the Bogoliubov decomposition
\begin{equation}\label{BD}
\hat{{\bar U}}(s,\tau)={\cal U}_{0}(\tau)+\hat{U}_{1}(s,\tau),
\end{equation}
where ${\cal U}_{0}(\tau)={\cal U}_{0}(0,\tau)$  is a classical SLDS background for $s=0$, $\hat{U}_{1}$ is the annihilation operator of photons representing the quantum fluctuations (perturbations) on the SLDS background ${\cal U}_{0}(\tau)$, satisfying the commutation relation $[\hat{U}_{1}(s,\tau),\hat{U}_{1}^{\dag}(s,\tau')]=\delta(\tau-\tau')$.
For the convenience of the following calculations, we introduce $\hat{w}\equiv \hat{U}_{1}/(\sqrt{{\cal A}g\cos\vartheta})$, which satisfies  $[{\hat w}(s,\sigma),{\hat w}^\dag(s,\sigma')]=\delta(\sigma-\sigma')$.

Substituting the Bogoliubov decomposition (\ref{BD}) into the reduced QNLS Eq. (\ref{QNLSE2}) and neglecting the high-order terms of $\hat{U}_{1}$, we obtain the equation for $\hat{w}$ and $\hat{w}^{\dag}$:
\begin{align}\label{twody}
i\frac{\partial}{\partial s}\,\,\binom{\hat{w}}{\hat{w}^{\dagger}}-{\cal A}^{2}g^{2}\cos^2\vartheta\,\hat{\cal L}\,\,\binom{\hat{w}}{\hat{w}^{\dagger}}=0,
\end{align}
where
$\hat{\cal L}$ is a linear matrix operator, defined by
\begin{align}\label{L}
\hat{\cal L}=\left(
\begin{matrix}
{\cal M}+2i\gamma\frac{\partial}{\partial\sigma} & 2{\cal N} \\
-2{\cal N}^{\ast} & -{\cal M}+2i\gamma\frac{\partial}{\partial\sigma}
\end{matrix}\right),
\end{align}
with ${\cal M}=-\frac{\partial}{\partial\sigma^2}+4 \tanh^2\sigma-2+2\gamma^2$, ${\cal N}=(\tanh\sigma+i\gamma)^{2}$, and $\gamma=\tan\vartheta$.

To solve Eq.~(\ref{twody}) for all possible quantum fluctuations, the key is to find the eigenmodes of the operator $\hat{\cal L}$ and constitute an orthogonal and complete set of them. The difficulty of success for this  depends on the property of $\hat{\cal L}$. Obviously, $\hat{\cal L}$ is {\it non-Hermitian}; its adjoint operator is given by
\begin{align}\label{Ldag}
\hat{\cal L}^{\dagger}=\sigma_3\hat{\cal L}\sigma_3=\left(
\begin{matrix}
{\cal M}+2i\gamma\frac{\partial}{\partial\sigma} & -2{\cal N} \\
2{\cal N}^{\ast} & -{\cal M}+2i\gamma\frac{\partial}{\partial\sigma}
\end{matrix}\right),
\end{align}
where $\sigma_3=\left(\begin{matrix}1 & 0\\0& -1\end{matrix}\right)$
is a Pauli matrix. Although $\hat{\cal L}$ is not Hermitian, i.e. $\hat{\cal L}^{\dagger}\neq \hat{\cal L}$, but it is {\it pseudo-Hermitian} due to the property $\hat{\cal L}^{\dagger}=\sigma_3\hat{\cal L}\sigma_3$. It is known that such a Pseudo-Hermitian operator possesses real eigenvalues, and it is possible to get the eigenmodes of $\hat{\cal L}$ and $\hat{\cal L}^{\dagger}$, which can be complete and bi-orthonormal in mutually dual function spaces of $\hat{\cal L}$ and $\hat{\cal L}^{\dagger}$~\cite{Vladimir2016,Ashida2020}.

To get the eigenmodes explicitly, we make the Bogoliubov transformation to expand the quantum fluctuations ${\hat w}$ as the form
\begin{eqnarray}\label{w}
{\hat w}(s,\sigma)
=&&\sum_{n}\left[u_{n}(\sigma)\hat{a}_{n}(s)+v^{\ast}_{n}(\sigma)\hat{a}_{n}^{\dagger}(s)\right]\nonumber\\
&&+\int dk\,\left[u_k(\sigma)\hat{a}_k(s)+v^{\ast}_k(\sigma)\hat{a}^{\dagger}_k(s)\right].
\end{eqnarray}
Here the indices $n$ and $k$ are quantum numbers denoting respectively the discrete and continuous modes;
$\hat{a}_{n}(s)$ and $\hat{a}_k(s)$ are respectively annihilation operators of photons for the discrete and continuous modes, satisfying respectively  the commutation relations $[\hat{a}_{n}(s),\hat{a}_{m}^\dag(s)]=\delta_{mn}$ and $[\hat{a}_k(s),\hat{a}^\dag_{k'}(s)]=\delta (k-k')$;
$u_{n}(\sigma)$, $v_{n}(\sigma)$, $u_k(\sigma)$, and $v_k(\sigma)$
are mode functions for the discrete and continuous spectra, respectively.

Assuming $\hat{a}_j (s)=\hat{a}_j (0)e^{-iE_j {\cal A}^2 g^2 \cos^2\vartheta\,s}$, and substituting it into Eq.~(\ref{twody}), we obtain the eigenvalue equations (i.e. BdG equations)
\begin{eqnarray}\label{BdGE}
& \hat{\cal L}\,|\Psi_{j} (\sigma)\rangle &=E_{j}|\Psi_{j}(\sigma)\rangle. \label{rel11}
\end{eqnarray}
Here, for simplicity, we have used the index $j$ to denotes both the discrete (for $j=n$) and continuous (for $j=k$) spectra. The eigen vectors $|\Psi_{j}(\sigma)\rangle\equiv (u_{j}(\sigma),v_{j}(\sigma))^T$, with symbol $``T$'' representing transpose.

Note that the eigenmodes of the operator $\hat{\cal L}$ forms a function space $V_L$, which is, however, not a Hilbert space because $\hat{\cal L}$ is not Hermitian. Following standard method~\cite{Faisal1981,SYLee2009,Brody2013,Vladimir2016,Ashida2020}, we consider the dual space  of $\hat{\cal L}$, i.e. the function space $V_{L^{\dag}}$ of the operator $\hat{\cal L}^{\dagger}$. The eigenvalue problem of $\hat{\cal L}^{\dagger}$ reads
\begin{eqnarray}
&\hat{\cal L}^{\dagger}|\Phi_{j} (\sigma)\rangle&=E_{j}^*|\Phi_{j}(\sigma)\rangle
=E_{j}|\Phi_{j}(\sigma)\rangle \label{rel12}
\end{eqnarray}
because $E_{j}$ is real. The above equation is usually written as the form
$\langle \Phi_{j} |\hat{\cal L}=E_{j} \langle \Phi_{j} |$~\cite{Vladimir2016}.
It is easy to show that $|\Phi_j(\sigma)\rangle =\sigma_{3}|\Psi_j(\sigma)\rangle$.

With the eigenmodes of the two mutually dual function spaces $V_L$ and $V_{L^{\dag}}$, we can define the product between the right vectors
$\{|\Psi_{j}\rangle\}$ and the left vectors $\{\langle \Phi_{j}|\}$:
\begin{align}\label{inner}
\langle\Phi_{j}|\Psi_{l}\rangle=\int_{-\infty}^{\infty}d\sigma
\langle \Phi_{j} (\sigma)|\Psi_{l}(\sigma)\rangle.
\end{align}
It can be shown that  $\{|\Psi_{j}\rangle\}$ and  $\{\langle \Phi_{j}|\}$ are bi-orthogonal and can be made to be normalized, i.e.
\begin{equation}
\langle\Phi_{j}(\sigma)|\Psi_{l}(\sigma)\rangle=\delta_{jl}.\label{orthonormality}
\end{equation}

To implement the perturbation calculation on the quantum fluctuations of the SLDS, one needs the eigenmodes in the dual spaces not only to be bi-orthonormalized, but also to span a complete set, so that any possible quantum fluctuations can be expanded as their linear superposition. However, the proof of the completeness for such bi-orthogonal eigenmodes is usually not an easy problem. If the all eigenmodes (especially the independent zero modes) are found, the completeness relationship reads
\begin{eqnarray}
&& \sum_{n}|\Phi_{n}
(\sigma)\rangle\langle\Psi_{n}(\sigma')|
+\int_{-\infty}^{+\infty}dk|\Phi_k(\sigma)\rangle\langle\Psi_k(\sigma')| \nonumber\\
&& =I\delta(\sigma-\sigma'),\label{completeness}
\end{eqnarray}
with $I$ the $2\times 2$ unit matrix.

\subsubsection{Eigenmode solutions and their bi-orthogonality and completeness}

The eigenmodes for the continuum spectrum (i.e. Goldstone bosons) of the present
eigenvalue problem [i.e. the BdG equations (\ref{BdGE})\,] can be found by the way similar to Refs.~\cite{HUANG1,YAN1}. We obtain the solutions
\begin{subequations}\label{ContinuousM}
\begin{eqnarray}
&& |\Psi_k(\sigma)\rangle=\binom{u_k(\sigma)}{v_k(\sigma)},\\
&& |\Phi_k(\sigma)\rangle =\sigma_{3}|\Psi_k(\sigma)\rangle=\binom{u_k(\sigma)}{-v_k(\sigma)},
\end{eqnarray}
\end{subequations}
where the eigenvalues and eigenfunctions given by
\begin{subequations}\label{ContinuousM1}
\begin{eqnarray}
&& E_{k}^{(\pm)}=|k|\left[-2\gamma\pm \sqrt{k^2+4(1+\gamma^2)}\right],\\
&& u_{k}(\sigma)=\frac{e^{i k \sigma}\left\{\tanh \sigma+\frac{i}{2}\left[{\cal D}_{\pm}(k)-k\right]\right\}^{2}}{\sqrt{2 \pi |k| \nu (k)} {\cal D}_{\pm}(k)},\\
&& v_{k}(\sigma)=\frac{e^{i k \sigma}\left\{\tanh \sigma-\frac{i}{2}\left[{\cal D}_{\pm}(k)+k\right]\right\}^{2}}{\sqrt{2 \pi |k| \nu (k)} {\cal D}_{\pm}(k)},
\end{eqnarray}
\end{subequations}
with  ${\cal D}_{\pm}(k)=-2\gamma\pm\nu(k)$ and $\nu(k)=\sqrt{k^2+4(1+\gamma^2)}$.
Note that the eigenmodes corresponding to the eigenvalue $E_{k}^{(-)}$ is non-physical and hence must be excluded. The reason is due to the fact $\lim_{ k\rightarrow\infty} E_{k}^{(-)}=-\infty$; with such an eigenvalue, the energy has no lower bound, which makes the system collapse.

The continuum eigenmode set $\{|\Psi_k(\sigma)\rangle\}$ and $\{|\Phi_k(\sigma)\rangle\}$ are not enough to constitute a complete bi-orthonormal basis. In fact, the operator $\hat{\cal L}$ and $\hat{\cal L}^{\dag}$ allow discrete eigenmodes with zero eigenvalues, i.e. zero modes, satisfying the equations
\begin{subequations}\label{dis0}
\begin{align}
\hat{\cal L}|\Psi_{n}(\sigma)\rangle=0,\,\,|\Psi_{n}(\sigma)\rangle&
=\binom{u_{n}(\sigma)}{v_{n}(\sigma)},\\
\hat{\cal L}^{\dagger}|\Phi_{n}(\sigma)\rangle=0,\,\,|\Phi_{n}(\sigma)\rangle&
=\sigma_{3}|\Psi_{n}(\sigma)\rangle,
\end{align}
\end{subequations}
$n=1,2,\cdots$.

Since a linear superposition of multiple zero modes is also a zero mode, in general one can get infinite many zero modes. How to choose these zero modes? The criterion for the choice of the zero modes is the following: (i)~They must be independent each other; (ii)~The set consisting of the continuous modes given in (\ref{ContinuousM}) and the zero modes (\ref{dis0}) should constitute a complete and bi-orthonormal basis, which is necessary for providing an expansion basis to express any quantum fluctuation of the soliton.

Based on such a criterion, we find that the system supports only single zero mode, given by
\begin{subequations}\label{dis1}
\begin{align}
& |\Psi_{1}(\sigma)\rangle =\binom{u_{1}(\sigma)}{v_{1}(\sigma)},\\
& |\Phi_{1}(\sigma)\rangle=\sigma_{3}|\Psi_{1}(\sigma)\rangle
=\binom{u_{1}(\sigma)}{-v_{1}(\sigma)},
\end{align}
\end{subequations}
with
\begin{subequations}\label{dis2}
\begin{align}
u_{1}(\sigma)=\frac{2{\rm sech}^2\sigma+i\gamma\tan\sigma+i\gamma\sigma{\rm sech}^{2}\sigma+1}{2\sqrt{2}},\\
v_{1}(\sigma)=\frac{2{\rm sech}^2\sigma+i\gamma\tan\sigma+i\gamma\sigma{\rm sech}^{2}\sigma-1}{2\sqrt{2}}.
\end{align}
\end{subequations}
It can be rigorously proved  that, for our present system,  the eigenmode set $\{\,|\Psi_{k}(\sigma)\rangle, |\Psi_{1}(\sigma)\rangle\,\}$ and $\{\,|\Phi_{k}(\sigma)\rangle, |\Phi_{1}(\sigma)\rangle\,\}$
given by (\ref{ContinuousM1}) (taking only the $E_{k}^{(+)}$-mode)
and (\ref{dis2})  is not only bi-orthonormal, but also complete, satisfying (\ref{orthonormality}) and {\ref{completeness}). A detailed proof for this is presented in Appendix~\ref{app3}. When $\gamma=0$ (i.e. for the special case of black soliton), these results are consistent with that obtained in previous studies~\cite{Yu2004,Huang2008}.

Taking the transformations
\begin{subequations}
\begin{align}
& u_{1}=\frac{1}{\sqrt{2}}\left(\psi_{1}+\phi_{1}\right),\\
& v_{1}^{\ast}=\frac{1}{\sqrt{2}}\left(\psi_{1}-\phi_{1}\right),\\
&\hat{Q}_{1}=\frac{1}{\sqrt{2}}\left(\hat{a}_{1}+\hat{a}_{1}^\dag\right),\\
&\hat{P}_{1}=\frac{1}{\sqrt{2}i}\left(\hat{a}_{1}-\hat{a}_{1}^\dag\right),
\end{align}
\end{subequations}
where $\hat{Q}_{1}$ and $\hat{P}_{1}$ are ``coordinate'' operators and ``momentum'' operators, satisfying the commutation relation $[\hat{Q}_{1},\hat{Q}_{1}]=[\hat{P}_{1},\hat{P}_{1}]=0$, and $[\hat{Q}_{1},\hat{P}_{1}]=i$,
we obtain the general expression of the quantum fluctuations of the SLDS, of the following form
\begin{eqnarray}\label{w1}
&& \hat{w}=\psi_{1}(\sigma)\hat{Q}_{1}(s)+i\phi_{1}(\sigma)\hat{P}_{1}(s),\nonumber\\
&& \hspace{0.6 cm}+\int_{-\infty}^{+\infty}dk\left[u_{k}(\sigma)\hat{a}_{k}(s)
+v^{\ast}_{k}(\sigma)\hat{a}_{k}^{\dagger}(s)\right].
\end{eqnarray}
Based on this expression, the effective Hamiltonian  (\ref{EffectiveH}) is diagonalized to be the form
\begin{align} \label{Dia-Hamiltonian}
\hat{H}_{\rm eff}&={\cal A}^{2}g^{2}\cos^2\vartheta \notag\\
&\hspace{0.5 cm}\times\left[\frac{1}{2}P_{1}^{2}(s)+
\int_{-\infty}^{+\infty}dkE_{k}^{(+)}\hat{a}_{k}^{\dagger}(s)\hat{a}_{k}(s)
\right],
\end{align}
with $\psi_{1}(\sigma)={\rm sech}^{2}\sigma$, $\phi_{1}(\sigma)=(i\gamma\tan\sigma+i\gamma\sigma{\rm sech}^{2}\sigma+1)/2$.   Fig.~\ref{Fig2}(a)-(d) shows the profiles of $\psi_{1}(\sigma)$, $\phi_{1}(\sigma)$, $u_k(\sigma)$, and $v_k(\sigma)$ as functions of $\sigma$, respectively. The term $P_1^2/2$ in (\ref{Dia-Hamiltonian}) is contributed by the zero mode.
We see that the zero mode behaves like a free particle and its mass is positive, which means that the SLDS is quite stable when the quantum fluctuations exist in the system.
\begin{figure}
\centering
\includegraphics[width=1\columnwidth]{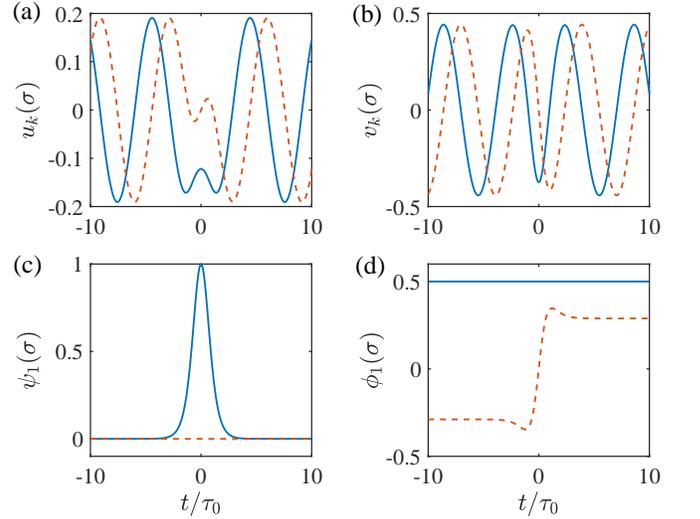}
\caption{~(a)-(d) Eigenmode functions $u_{k}(\sigma)$, $v_{k}(\sigma)$, $\psi_{1}(\sigma)$, and $\phi_{1}(\sigma)$, respectively, plotted with wavenumber k=1 and blackness parameter $\vartheta=\pi/6$. The solid (dashed) line represents the relevant real (imaginary) part.}
\label{Fig2}
\end{figure}

We must stress that the zero mode given by (\ref{dis1}) is not corresponding to a Goldstone boson, though its origin has some similarities to that of Goldstone bosons~\cite{note3}. The zero mode is a discrete eigenmode of the system. It is different from the Goldstone bosons described by the continuous eigenmodes, which also have a zero frequency as its lower bound of the continuum of frequency. In fact, the appearance of the zero mode is due to the existence of the SLDS, which is inhomogeneous in space. If the classical background ${\cal U}_0$ in the Bogoliubov decomposition (\ref{BD}) is a constant, the zero mode will disappear. For a similar discussion in relativistic quantum field theory, see Ref.~\cite{Raj1987}.

\section{Quantum squeezing of slow-light dark solitons}\label{Sec4}

\subsection{Quantum dynamics of slow-light dark solitons}\label{Sec4a}

Based on the diagonalized effective Hamiltonian (\ref{Dia-Hamiltonian}), it is easy to study the quantum dynamics of the SLDS. The Heisenberg equations of motion
for $\hat{Q}_{1}(s)$, $\hat{P}_{1}(s)$, and $\hat{a}_{k}(s)$ read
\begin{subequations}\label{HE0}
\begin{align}
& \frac{\partial}{\partial s}\hat{Q}_{1}(s)-{\cal A}^{2}g^{2}\cos^2\vartheta\hat{P}_{1}(s)=0, \\
& \frac{\partial}{\partial s}\hat{P}_{1}(s)=0,  \\
& i\frac{\partial}{\partial s}\hat{a}_{k}(s)-{\cal A}^{2}g^{2}\cos^2\vartheta E_{k}^{(+)}\hat{a}_{k}(s)=0.
\end{align}
\end{subequations}
The exact solutions of these equations can be obtained, given by
\begin{subequations}\label{SolutionofHE}
\begin{align}
& \hat{Q}_{1}(s)=\hat{Q}_{1}(0)+{\cal A}^{2}g^{2}\cos^2\vartheta\hat{P}_{1}(0)s, \\
& \hat{P}_{1}(s)=\hat{P}_{1}(0),\\
& \hat{a}_{k}(s)=\hat{a}_{k}(0)e^{-i{\cal A}^{2}g^{2}\cos^2\vartheta E_{k}^{(+)        }s},
\end{align}
\end{subequations}
where $\hat{Q}_{1}(0)$, $\hat{P}_{1}(0)$, $\hat{a}_{k}(0)$ are the values of $\hat{Q}_{1}(s)$, $\hat{P}_{1}(s)$, $\hat{a}_{k}(s)$ at $s=0$, respectively.
From Eqs.~(\ref{HE0}) and their solutions (\ref{SolutionofHE}), we have the following conclusions: (i)~The quantum fluctuations contributed by the zero mode displays specific characters. The momentum operator $\hat{P}_{1}$ remains unchanged during propagation (i.e. the momentum of the SLDS is conserved); while the evolution of the position operator $\hat{Q}_{1}$ depends on $\hat{P}_{1}(0)$, the value of the momentum operator  $\hat{P}_{1}$ at $s=0$. Such a correlation between $\hat{Q}_{1}$ and $\hat{P}_{1}$ leads to a position spreading of the SLDS, contributed by the Kerr nonlinearity (characterized by the nonlinear parameter $g$). However, photon-number and phase fluctuations are not predicted here, different from the case of bright solitons~\cite{Zhu2021,Yan1998PRE,Huang2006}.
(ii)~The quantum fluctuation of the continuum modes (characterized by the quantum number $k$) has only a simple effect, i.e. a phase shift to the same mode caused by the Kerr nonlinearity.

From the formulas (\ref{CS0}) and (\ref{w}), we can get the approximate expression of the quantized probe field by using renormalization technique~\cite{Nayeh}, given by
\begin{align}\label{ueqs}
\hat{\bar{U}}\left(s,\sigma\right)
\approx{\cal A}\sqrt{g}\left[\cos\vartheta\tanh(\sigma+\frac{\hat{Q}_{1}}{\sqrt{g}})
+i\sin\vartheta\right]e^{i\theta_0+i\frac{\hat{P}_{1}\sigma}{{\cal A}\sqrt{g}}}.
\end{align}
One can see clearly that the quantum fluctuations of the SLDS are mainly contributed by the zero mode, which propagates together with the soliton; the conjugated operator pair $\hat{Q}_{1}$ and $\hat{P}_{1}$ describe the position and momentum fluctuations, respectively. The reason for no fluctuation of particle number (amplitude) and phase is as follows. In our approach, the SLDS has an infinite large background [i.e. it contains very large (infinite) photon number], and hence no phase diffusion occurs in the presence of perturbations. If,  however, the system has a finite size (e.g. when there is an external potential $V_{\rm ext}$ acting on the system, the photon number in the soliton will be finite), a phase diffusion of the soliton will happen.

With these results, we can give a numerical estimation on the quantum fluctuations of the SLDS. Let $|\Psi\rangle =|n_{0},n_1,n_c\rangle$
denotes the quantum state with $n_{0}$ photons in the SLDS; $n_1$
photons in the zero mode, and $n_c$ photons in the continuous modes. We assume that, at the entrance of the system ($s=0$), the quantum state of the probe field is in the ``vacuum'' state $|\Psi_0 \rangle=|n_{0},0,0\rangle$ (i.e. the probe field has no quantum fluctuation). Based on the analytical result (\ref{SolutionofHE}),
we obtain $\langle \hat{Q}_{1}(s)\rangle=\langle \hat{P}_{1}(s)\rangle=0$,
$\langle \hat{Q}_{1}^2(0)\rangle=\langle \hat{P}_{1}^2(0)\rangle=1/2$;
the variances (mean-squared derivations) as functions of $s$ are given by
\begin{align}\label{unc}
 &\langle\hat{P}_{1}^2(s)\rangle=\frac{1}{2},\\
 &\langle\hat{Q}_{1}^2(s)\rangle=\frac{1}{2}\left(1+{\cal A}^4g^4\cos^4\vartheta s^2\right),
\end{align}
here $\langle\cdots\rangle \equiv\langle\Psi_0|\cdots|\Psi_0\rangle$.  One sees that the variance of position fluctuation is propagation dependent, while the variance of the momentum fluctuation is a constant during propagation.

\subsection{Quantum squeezing of slow-light dark solitons}\label{sec4b}

In recent years, a multitude of studies have been paid to quantum squeezing~\cite{Ma2011,Andersen2016}. In particular, many efforts have focused on the quantum squeezing of light, which has important applications, especially for quantum precision measurements (e.g. the detection of gravitational waves)~\cite{Schnabel2017}. The results obtained above can be exploited to investigate the quantum squeezing of the SLDS, which can be measured using a homodyne detection method~\cite{HausJOSAB1990,YLai1993}.
In comparison with the zero mode, the quantum fluctuations from the continuous modes are much weaker and hence will be neglected in the following calculation.

The quantum squeezing of the SLDS may be described by the quadrature operators at the angle $\theta$ related to the operator $\hat{a}_{1}$~\cite{Mandel}
\begin{eqnarray} \label{QuadOperators}
\hat{X}_{\theta}(s)
&& =\frac{1}{\sqrt{2}}\left[\hat{a}_{1}(s)\,e^{-i\theta}
+\hat{a}_{1}^{\dag}(s)\,e^{i\theta}\right]\label{QuadOperators1}\nonumber\\
&& =\hat{Q}_{1}(s)\cos\theta+\hat{P}_{1}(s)\sin\theta,\label{QuadOperators2}
\end{eqnarray}
which satisfies the commutation relation $[{\hat X}_{\theta},{\hat X}_{\theta+\frac{\pi}{2}}]=i$.
With the results obtained in the last subsection, it is easy to get the expression of the variance of $\hat{X}_{\theta}$:
\begin{align}\label{XVariance}
&\langle \hat{X}_{\theta}^2(s) \rangle =\frac{1}{2}\left({\cal A}^2 g^2 s\,\cos^2\vartheta \cos\theta +\sin\theta\right)^2+\frac{\cos^2\theta}{2},
\end{align}

Shown in  Fig.~\ref{Fig3}(a)
\begin{figure}
\centering
\includegraphics[width=1\columnwidth]{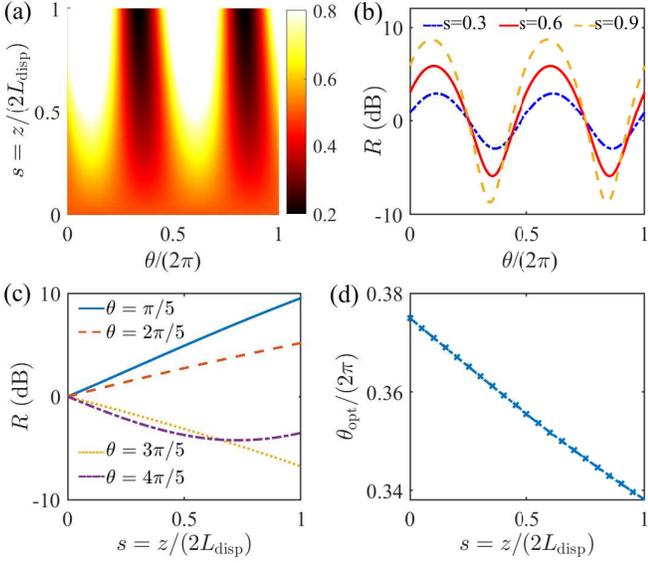}
\caption{(a)~Quadrature variance $\langle\hat{X}_{\theta}^2\rangle$  of the SLDS as a function of $s=z/(2L_{\rm disp})$ (dispersion length $L_{\rm disp}$=0.95\,cm) and $\theta/(2\pi)$.
Different colors shown in the color bar  denote different magnitudes of $\langle\hat{X}_{\theta}^2\rangle$. The the quadrature variance in the black domains is much smaller than its vacuum value, indicating that the SLDS displays larger quadrature squeezing.
(b)~Squeezing ratio $R~({\rm unit\,\, dB})$ versus detection angle $\theta$ with $s =0.3, 0.6 , 0.9$, plotted by dotted blue, solid red, and dashed yellow lines, respectively.
(c)~$R$ versus propagation distance $s$ with $\theta =\pi/5, 2\pi/5 , 3\pi/5, 4\pi/5$.
(d)~Optimum angle $\theta_{\rm opt}$ for the quadrature variance $\langle\hat{X}_{\theta}^2\rangle$ as a function of propagation distance $s$. (a)-(d) are all plotted for ${\cal A}=1$, $\vartheta=0$, and $g=1$.
}
\label{Fig3}
\end{figure}
is  $\langle\hat{X}_{\theta}^2\rangle$  as a function of $s=z/(2L_{\rm disp})$ and $\theta/(2\pi)$ by taking ${\cal A}=1$, $\vartheta=0$, and $g=1$. We see that when $s=0$, the variance takes the vacuum value  $\langle\hat{X}_{\theta}^2(0)\rangle=1/2$; for any $s$, $\langle\hat{X}_{\frac{\pi}{2}}^2(s)\rangle=1/2$.
However, when $\theta$ and $s$ locate in the black domains of the figure, the quadrature variance is much smaller than its vacuum value, which means that the SLDS can be significantly quantum-mechanically squeezed. The SLDS can also be made to be anti-squeezed, which occurs in the bright domains of the figure.

One can also define the squeezing ratio, i.e. the ratio of the quadrature variance between the value at position $s$ and that at the position $s=0$~\cite{HausJOSAB1990,YLai1993}
\begin{align}\label{R}
R=\frac{\langle\hat{X}_{\theta}^2(s)\rangle}{\langle\hat{X}_{\theta}^2(0)\rangle},
\end{align}
to characterize the degree of squeezing quantitatively.

Fig.~\ref{Fig3}(b) shows the squeezing ratio $R$  (with unit dB) of the SLDS as a function of angle $\theta$ for different propagation distance $s=0.3$, 0.6, and 0.9, respectively.
We see that the squeezing ratio is sensitive to the selections of $\theta$. Illustrated in Fig.~\ref{Fig3}(c) is the degree of squeezing (also antisqueezing) in the system, which becomes larger during propagation (i.e. when $s$ increases). However, at some special detection angle (such as $4\pi/5$), the squeezing reaches a threshold.
We stress that, in comparison with the quantum squeezing of dark solitons in optical fibers, the quantum squeezing of the SLDS in the present atomic gas is more significant. The typical feature is that the SLDS can acquire a large quantum squeezing in a very short propagation distance (in the order of centimeter). The physical reason is that the EIT-based atomic gas possesses larger Kerr nonlinearity (much bigger than that in optical fibers), which makes the typical nonlinearity length $L_{\rm nonl}$ of the system be very small (for the case of the SLDS in the present system, one has $L_{\rm nonl}\approx L_{\rm disp}=0.95$\,cm). In addition, the ultraslow propagating velocity of the SLDS is another factor that makes the soliton squeezing more efficient.

By minimizing the quadrature variance $\langle\hat{X}_{\theta}^2\rangle$
[Eq.~(\ref{XVariance})] with respect to $\theta$, we can obtain the optimum angle as a function of the propagation distance $s$, i.e. $\theta_{\rm opt}=\theta_{\rm opt} (s)$, which is plotted in Fig.~\ref{Fig3}(d).
Once $\theta_{\rm opt}(s)$ is known, experimentally one can choose the optimum detection angle to acquire the largest suppression of the quantum uncertainties in the position and momentum of the SLDS. With $\theta_{\rm opt} (s)$ we can get the minimum value of the quadrature as a function of $s$; meanwhile, the quadrature for the angle $\theta_{\mathrm{opt}}+\pi/2$  will be maximized.

Shown in Fig.~\ref{Fig4}(a)
\begin{figure}
\centering
\includegraphics[width=1\columnwidth]{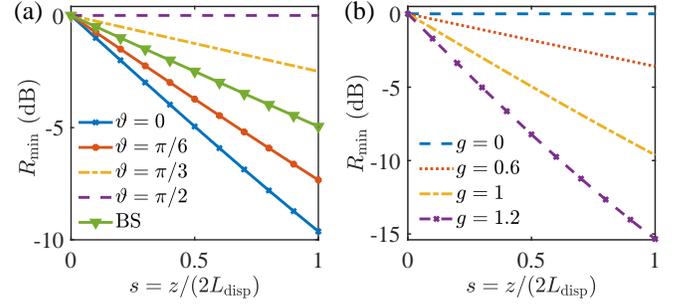}
\caption{
(a)~Minimum squeezing ratio $R_{\rm min}$ as a function of $s=z/(2L_{\rm disp})$ and the blackness parameter  $\vartheta=0$, $\pi/3$, $\pi/2$, and $\pi/6$ (with ${\cal A}=1$, $g=1$). ``BS'' means the result for slow-light bright soliton.
(b)~$R_{\rm min}$ as a function of $s$ versus the nonlinear coefficient $g=0$, $g=0.6$, $g=1$, $g=1.2$
(with ${\cal A}=1$, $\vartheta=0$).
}
\label{Fig4}
\end{figure}
is the minimum squeezing ratio $R_{\rm min}$ as a function of $s=z/(2L_{\rm disp})$ and the blackness parameter  $\vartheta=0$, $\pi/3$, $\pi/2$, and $\pi/6$, for ${\cal A}=1$ and $g=1$. One sees that: (i)~$R_{\rm min}$ is lowered as $s$ is increased; (ii)~$R_{\rm min}$ is strongly dependent on the parameter $\vartheta$ (which characterizing the blackness of the SLDS; see Sec.~\ref{section3a}). The darker the SLDS, the larger the minimum squeezing ratio $R_{\rm min}$. In the figure, the  $R_{\rm min}$ for slow-light bright soliton is also plotted, given by the green line with triangles. We see that if the SLDS is made to be dark enough (i.e.  $\vartheta$ is small), its minimum squeezing ratio $R_{\rm min}$ can be much smaller than the slow-light bright soliton. This means that the SLDS can have large quantum squeezing than that of the slow-light bright soliton.

We stress that the minimum squeezing ratio $R_{\rm min}$ is strongly dependent on the Kerr nonlinearity of the system, which is proportional to the soliton's amplitude.  Fig.~\ref{Fig4}(b) shows the result of $R_{\rm min}$  as a function of $s$ for the nonlinear coefficient $g=0$, $g=0.6$, $g=1$, $g=1.2$ (with ${\cal A}=1$, $\vartheta=0$). We see that $R_{\rm min}$ decreases rapidly as
$g$ increases. Because the EIT effect can result in a large enhancement of the Kerr nonlinearity and the Kerr nonlinearity can be actively controlled due to the active character of the system (e.g., the nonlinear parameter $g$ can be adjusted by changing the two-photon detuning $\Delta_2$), the EIT-based atomic gas is an excellent platform for realizing large quantum  squeezing of the SLDS.

Finally, we indicate that the larger Kerr  nonlinearity contributed by the EIT  can not only result in the large quantum squeezing of the probe laser pulse (here the SLDS), but also induce an atomic spin squeezing in the system.  The result of the atomic spin squeezing in the presence of the SLDS is given in Appendix~\ref{app4}.

\section{Summary}\label{Sec5}

In this work, we have investigated the quantum effect of SLDS in a cold atomic gas with defocuing Kerr nonlinearity working on the condition of EIT. We have made an analytical calculation on the quantum fluctuations of the SLDS through solving the BdG equations and the relevant non-Hermitian eigenvalue problem. We found that only a single zero mode is allowed for the quantum fluctuations, which is different from the quantum fluctuations of bright solitons where two independent zero modes occur. We have rigorously proved that the eigenmodes, which consist of continuous modes and the zero mode, are bi-orthogonal and constitute a complete bi-orthonormalized basis, which is useful and necessary for the calculation of the quantum fluctuations of the SLDS. We have demonstrated that, due to the large Kerr nonlinearity contributed from the EIT effect, a significant quantum squeezing of the SLDS can be realized; the squeezing efficiency can be manipulated by the Kerr nonlinearity and the blackness and amplitude of the soliton, which can be much higher than that of slow-light bright solitons. Our work is useful for developing quantum nonlinear optics and non-Hermitian Physics, and for applications in Bose-condensed quantum gases, quantum and nonlinear fiber optics, quantum information processing, and precision measurements, and so on.

\section*{Acknowledgments}
This work was supported by the National Natural Science Foundation of China under Grant No.~11975098.

\appendix

\section{Explicit expressions of the Heisenberg-Langevin equations}\label{app1}
Explicit expressions of the Heisenberg-Langevin equations (\ref{HLM}a) are given by
\begin{subequations}\label{HLMEs_explicit}
\begin{align}
& i\frac{\partial}{\partial t}{\hat S}_{22}-i\Gamma_{23}{\hat S}_{33}-\Omega_{c}{\hat S}_{23}+\Omega_{c}^{\ast}{\hat S}_{32}-i{\hat F}_{22}=0,\\
& i\left(\frac{\partial}{\partial t}+\Gamma_{3}\right){\hat S}_{33}+g_{p}{\hat S}_{13}\hat{E}_{p}-g_{p}^{\ast}\hat{E}_{p}^{\dag}{\hat S}_{31}+\Omega_{c}{\hat S}_{23}\nonumber\\
& -\Omega_{c}^{\ast}{\hat S}_{32}-i{\hat F}_{33}=0,\\
& \left(i\frac{\partial}{\partial t}+d_{21}\right){\hat S}_{21}+\Omega_{c}^{\ast}{\hat S}_{31}-g_{p}{\hat S}_{23}\hat{E}_{p}-i{\hat F}_{21}=0,\\
& \left(i\frac{\partial}{\partial t}+d_{31}\right){\hat S}_{31}+\Omega_{c}{\hat S}_{21}+g_{p}({\hat I}-{\hat S}_{22}-2{\hat S}_{33})\hat{E}_{p}\nonumber\\
&-i{\hat F}_{31}=0,\\
& \left(i\frac{\partial}{\partial t}+d_{32}\right){\hat S}_{32}+\Omega_{c}\left({\hat S}_{22}-{\hat S}_{33}\right)+g_{p}{\hat S}_{12}\hat{E}_{p}\nonumber\\
&-i{\hat F}_{32}=0.
\end{align}
\end{subequations}
Here ${\hat S}_{11}={\hat I}-{\hat S}_{22}-{\hat S}_{33}$, ${\hat I}$ is identity operator, $d_{\alpha\beta}=\Delta_{\alpha}-\Delta_{\beta}+i\gamma_{\alpha\beta}$
($\alpha\neq \beta)$, $\gamma_{\alpha\beta}\equiv(\Gamma_\alpha+\Gamma_\beta)/2+\gamma_{\alpha\beta}^{\rm dep}$, $\Gamma_\beta\equiv\sum_{\alpha<\beta}\Gamma_{\alpha\beta}$, and
$\gamma_{\alpha\beta}^{\rm dep}$ is the dephasing rate between $|\alpha\rangle$ and $|\beta\rangle$.
${\hat F}_{\alpha\beta}$ are $\delta$-correlated Langevin noise operators associated with the dissipation in the system, with the two-time correlation function given by
\begin{align}\label{FF}
\langle\hat{F}_{\alpha\beta}(z,t)\,\hat{F}_{\alpha'\beta'}&(z',t')\rangle=\notag\\
&\frac{L}{N}\delta(z-z')\delta(t-t'){\cal D}_{\alpha\beta,\alpha'\beta'}(z,t),
\end{align}
where ${\cal D}_{\alpha\beta,\alpha'\beta'}$ is atomic diffusion coefficient~\cite{Kolchin}, which can be obtained from the Eqs.~(\ref{HLMEs_explicit}) using the generalized fluctuation dissipation theorem. Some of them are given by
\begin{subequations}
\begin{align}
{\cal D}_{21,12}&=\Gamma_{23}\langle\hat{S}_{33}\rangle,\\
{\cal D}_{31,13}&=0,\\
{\cal D}_{\alpha1,1\beta}&=0,
\end{align}
\end{subequations}
with $\alpha,\beta=2,3$\, ($\alpha\neq\beta$).

\section{Explicit expressions of $K(\omega)$,  $W$, and ${\hat{\cal F}}_{p}(z,t)$}\label{app2}

The linear dispersion relation reads
\begin{equation}\label{H}
K(\omega)=\frac{\omega}{c}+\frac{|g_{p}|^{2}N}{c}\frac{\omega+d_{21}}{D(\omega)},
\end{equation}
Here $\omega$ is the sideband frequency of the probe pulse. The new noise operator $\hat{{\cal F}}_{p}(z,t)$ is defined by
\begin{equation}
\hat{{\cal F}}_{p}(z,t)=\frac{ g_{p}^\ast N}{c}\frac{\left(\omega+d_{21}\right) \hat{F}_{31}(z,t)-\Omega_{c}\hat{F}_{21}(z,t)}{D(\omega)},
\end{equation}
with $D(\omega)=|\Omega_{c}|^2-(\omega+d_{21})(\omega+d_{31})$.

By considering the steady-state solution  of the Heisenberg-Langevin equations (for which ${\hat S}_{11}={\hat I}$ and ${\hat S}_{22}={\hat S}_{33}=0$), we obtain the solution at the first-order approximation, given by
\begin{equation}\label{SE}
{\hat S}_{\alpha1}=a_{\alpha1}^{(1)}g_{p}\hat{E}_{p}\,(\alpha=2,3)
\end{equation}
and other ${\hat S}_{\alpha\beta}=0$, where
\begin{equation}\label{it1}
a_{\alpha1}^{(1)}=\frac{-\Omega_{c}^{\ast}\delta_{\alpha2}+d_{21}
\delta_{\alpha3}}{|\Omega_{c}|^2-d_{21}d_{31}}.
\end{equation}

Proceeding to the next order of iteration by substituting Eq.~(\ref{it1}) into
Eqs.~(\ref{HLMEs_explicit}), one obtains ${\hat S}_{\alpha\beta}=a_{\alpha\beta}^{(2)}|g_{p}|^2\hat{E}_{p}^{\dag}\hat{E}_{p}\,
(\alpha,\beta=1,2,3)$. Here
\begin{subequations}\label{it2}
\begin{align}
& a_{11}^{(2)}=\frac{\Gamma_{23}+2D_{c}}{\Gamma_{13}D_{c}}2{\rm Im}\left[a_{31}^{(1)\ast}\right]-\frac{1}{D_{c}}2{\rm Im}\left[\frac{\Omega_{c}^{\ast}}{d_{32}}a_{21}^{(1)\ast}\right],\\
& a_{22}^{(2)}=\frac{1}{D_{c}}2{\rm Im}\left[\frac{\Omega_{c}^{\ast}}{d_{32}}a_{21}^{(1)\ast}\right]
-\frac{\Gamma_{23}+D_{c}}{\Gamma_{13}D_{c}}2{\rm Im}\left[a_{31}^{(1)\ast}\right],\\
& a_{33}^{(2)}=-\frac{1}{\Gamma_{13}}2{\rm Im}\left[a_{31}^{(1)\ast}\right],\\
& a_{32}^{(2)}=-\frac{1}{d_{32}}\left[a_{21}^{(1)\ast}+\Omega_{c}
\left(a_{22}^{(2)}-a_{33}^{(2)}\right)\right],
\end{align}
\end{subequations}
and other ${\hat S}_{\alpha\beta}=0$, with $D_{c}=2\gamma_{32}|\Omega_{c}|^2/|d_{32}|^2$.

Based on the above results, we can proceed to the third-order of iteration. We get  
\begin{subequations}
\begin{align}\label{it3}
& {\hat S}_{31}=a_{31}^{(3)}|g_{p}|^2g_{p}\hat{E}_{p}^{\dag}\hat{E}_{p}\hat{E}_{p},\\
& a_{31}^{(3)}\equiv\frac{\Omega_{c}a_{32}^{(2)\ast}-d_{21}\left[a_{22}^{(2)}
+2a_{33}^{(2)}\right]}{|\Omega_{c}|^2-d_{21}d_{31}}.
\end{align}
\end{subequations}
The solutions of other ${\hat S}_{\alpha\beta}$ are also obtained but are omitted here.

{\color{blue}
The optical susceptibility of the probe field is defined by $\chi_{p}={\cal N}_{a}|{\bf e}_{p}\cdot\mathbf{p}_{31}|\,\rho_{31}/(\varepsilon_{0}{\cal E}_{p})$, with $\rho_{31}\equiv \langle\hat{S}_{31}\rangle$.
Based on the above result, we obtain $\chi_{p}=\chi_{p}^{(1)}+\chi_{p}^{(3)}|{\cal E}_{p}|^{2}$, with
the third-order Kerr nonlinear susceptibility given by
\begin{subequations}
\begin{align}\label{chi3}
& \chi_{p}^{(3)}=\frac{{\cal N}_{a}|\mathbf{p}_{31}|^{4}}{\varepsilon_{0}\hbar^{3}}a_{31}^{(3)}=\frac{2c|{\bf e}_{p}\cdot\mathbf{p}_{31}|^{2}}{\hbar^2\omega_{p}}W,\\
& W=\frac{{\cal N}_{a}|{\bf e}_{p}\cdot\mathbf{p}_{31}|^{2}\omega_{p}}{2c\varepsilon_{0}\hbar}a_{31}^{(3)}.
\end{align}
\end{subequations}
Generally, $\chi_{p}^{(3)}$ and $W$ are functions of $\omega$ (the sideband frequency of the probe pulse). Since we are interested in the probe-pulse propagation near the center frequency $\omega_p$, the coefficients in the QNLS equation (\ref{QNLS0}) will be estimated at $\omega=0$. In this case, these coefficients are functions of the one- and two-photon detunings (i.e. $\Delta_3$ and $\Delta_2$), and other system parameters.
}

\section{Proof on the completeness and bi-orthonormality of the eigenmode set}\label{app3}

For investigating the physical properties of the quantum fluctuations of the SLDS, it is necessary to acquire the all eigenmodes of the BdG eigenvalue problem (\ref{rel11}). In addition, the eigenmode set obtained should be complete and bi-orthonormal, which is necessary not only for a complete and correct description of quantum fluctuations of the SLDS, but also for obtaining a general and consistent perturbation expansion valid for any perturbation on the soliton when external and/or initial disturbances are applied into the system.

\subsubsection{Bi-orthogonality}

We first prove the bi-orthogonality for the continuous modes given by (\ref{ContinuousM}) and (\ref{ContinuousM1}).  Consider the following integral
\begin{align}\label{ortt}
&\hspace{0.2 cm}\langle\Phi_{k'}(\sigma)|\Psi_{k}(\sigma)\rangle\notag\\
&=\int_{-\infty}^{\infty}d\sigma\left[u_{k}(\sigma)u_{k'}^{\ast}(\sigma)-v_{k}^{\ast}(\sigma)v_{k'}(\sigma)\right]\notag\\
&=\int_{-\infty}^{\infty}d\sigma\frac{e^{i(k-k^{\prime}) \sigma}}{2 \pi \sqrt{k k^{\prime} \nu(k) \nu(k')}{\cal D}(k) {\cal D}(k^\prime)}\{A+B\},
\end{align}
with
\begin{align}
A&=\left\{-\frac{1}{4}\left[{\cal D}(k^\prime)-k^{\prime}\right]^{2}+1\right\}\left\{-\frac{1}{4}\left[{\cal D}(k)-k\right]^{2}+1\right\} \notag\\
&\hspace{2mm}+\left\{-\frac{1}{4}\left[{\cal D}(k^\prime)+k^{\prime}\right]^{2}+1\right\}\left\{\frac{1}{4}\left[{\cal D}(k)+k\right]^{2}-1\right\} \notag\\
&\hspace{2mm}-2 k {\cal D}(k^\prime)-2 k^{\prime} {\cal D}(k),\\
B&=i\left\{\left[{\cal D}(k)-{\cal D}(k^\prime)\right]\left[\frac{1}{2} {\cal D}(k) {\cal D}(k^\prime)+k k^{\prime}+2\right]\right.\notag\\
&\left.\hspace{0.8 cm}+\frac{1}{2}\left[k^{2} {\cal D}(k^\prime)-k^{\prime 2} {\cal D}(k)\right]\right\} \tanh \sigma\notag\\
&\hspace{2mm}+\left\{\left(2 k^{\prime}-k\right) {\cal D}(k)+\left(2 k-k^{\prime}\right) {\cal D}(k^\prime)\right\} \operatorname{sech}^{2} \sigma\notag\\
&\hspace{2mm}+2 i\left[{\cal D}(k^\prime)-{\cal D}(k)\right] \tanh \sigma \operatorname{sech}^{2} \sigma.
\end{align}
It is easy to show that the first term of the right hand side of Eq.~(\ref{ortt}) (related to $A$) equals to $\delta(k-k^{\prime})$. The second term (related to $B$) can be calculated by using the following formulae
\begin{subequations}
\begin{align}
&\int_{-\infty}^{\infty} e^{i k x} \tanh x d x=i \pi  {\rm csch} (\pi k / 2), \\
&\int_{-\infty}^{\infty} e^{i k x} \operatorname{sech}^{2} x d x=\pi k  {\rm csch} (\pi k / 2), \\
&\int_{-\infty}^{\infty} e^{i k x} \tanh x \operatorname{sech}^{2} x d x=i \pi k^{2}  {\rm csch} (\pi k / 2)/2.
\end{align}
\end{subequations}
It is also easy to show that the second term equals zero. Thereby,  we have
\begin{equation}
\langle\Phi_{k'}(\sigma)|\Psi_{k}(\sigma)\rangle=\delta(k-k').
\end{equation}

In a similar way, we can prove the bi-orthogonality of the zero mode
\begin{align}\label{orttdisc}
\langle\Phi_{1}(\sigma)|\Psi_{1}(\sigma)\rangle=1,
\end{align}
as well as the bi-orthogonality of between the zero mode and the continuous modes
\begin{eqnarray}
&&\langle\Phi_{1}(\sigma)|\Psi_{k}(\sigma)\rangle=0.
\end{eqnarray}

\subsubsection{Completeness}

To prove the completeness of the eigenmodes, we consider the expression $|\Phi_{1}(\sigma)\rangle\langle\Psi_{1}(\sigma')|
+\int_{-\infty}^{+\infty}dk|\Phi_k(\sigma)\rangle\langle\Psi_k(\sigma')|$.
Using the results (\ref{ContinuousM})- (\ref{dis2}), we have
\begin{align}\label{int1}
& |\Phi_{1}(\sigma)\rangle\langle\Psi_{1}(\sigma')|+\int_{-\infty}^{+\infty}dk
|\Phi_k(\sigma)\rangle\langle\Psi_k(\sigma')|
\notag\\
&=\left(
\begin{array}{cc}
 {\cal G}(\sigma)& {\cal R}(\sigma) \\
  {\cal R}(\sigma') & {\cal G}(\sigma')
\end{array}\right).
\end{align}
Here ${\cal G}(\sigma)=X(k)+\psi_{1}(\sigma)\phi_{1}^{\ast}(\sigma')+\psi_{1}^{\ast}(\sigma)\phi_{1}(\sigma')$, ${\cal R}(\sigma)=Y(k)+\psi_{1}(\sigma)\phi_{1}(\sigma')-\phi_{1}(\sigma)\psi_{1}(\sigma')$; $X(k), Y(k)$ are given by
\begin{subequations}
\begin{align}
X(k)&=\int_{-\infty}^{+\infty}\left[u_{k}(\sigma)u_{k}^{\ast}(\sigma')-v_{k}^{\ast}(\sigma)v_{k}(\sigma')
\right]dk,\label{int2}\\
Y(k)&=\int_{-\infty}^{+\infty}\left[u_{k}(\sigma)v_{k}^{\ast}(\sigma')-v_{k}^{\ast}(\sigma)u_{k}(\sigma')\right]dk.
\label{int3}
\end{align}
\end{subequations}

Functions $X(k)$ and $Y(k)$ have the following properties:
(i)~As $k\rightarrow\infty$, $X(k) \rightarrow\delta(\sigma-\sigma')$, $Y(k)\rightarrow0$; (ii)~$X(k)$ and $Y(k)$ are analytic in the complex $k$ plane, except for the existence of a single pole at $k = 0$, two second order poles at $k = k_P = \pm2i$, and two branch points at $k = k_b = \pm2i\sqrt{1+\gamma^2}$. For convenience, we introduce $P(k)\equiv X(k)-\delta(\sigma-\sigma')$. According to Jordan’s lemma, we can use the residue theorem to calculate the integrals in (\ref{int2}) and (\ref{int3}), with the integral path shown in Fig.~\ref{Fig5}.
\begin{figure}
\centering
\includegraphics[width=1\columnwidth]{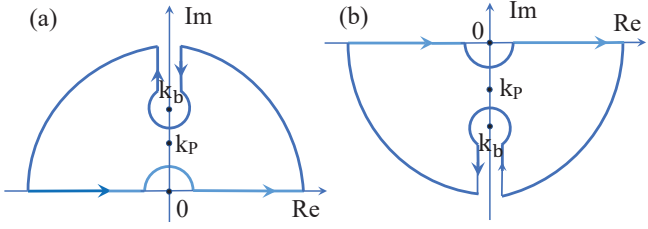}
\caption{(a)~Integral path of P(k) for $\sigma-\sigma^{\prime}>0$, with $k_{P}=2 i, k_{b}=2 i \sqrt{1+\gamma^{2}}$.
(b)~The same as (a) but for $\sigma-\sigma^{\prime}<0$,  with $k_{P}=-2 i, k_{b}=-2 i \sqrt{1+\gamma^{2}}$. Here Re and Im are the real and imaginary part of complex $k$, respectively.
}
\label{Fig5}
\end{figure}

After a detailed calculation, we obtain\, [with Res[$P(k_s)$] representing the reside of $P(k)$ at $k=k_s$]
\begin{subequations}
\begin{align}\label{comp}
P(k)&=\begin{cases}2\pi i{\rm Res}[P(2i)]+\pi i{\rm Res}[P(0)], \hspace{0.54 cm}  \text { for }\hspace{0.01 cm}  \sigma-\sigma'>0 \\
-2\pi i{\rm Res}[P(-2i)]-\pi i{\rm Res}[P(0)], \hspace{0.01 cm}\text { for }\hspace{0.01 cm} \sigma-\sigma'<0\end{cases}\notag\\
&=-\frac{1}{2}{\rm sech}^2\sigma\left(-i\gamma\tan\sigma'-i\gamma\sigma'{\rm sech}^{2}\sigma'+1\right)\notag\\
&\hspace{1 cm}-\frac{1}{2}{\rm sech}^2\sigma'\left(i\gamma\tan\sigma+i\gamma\sigma{\rm sech}^{2}\sigma+1\right),\notag\\
&=-\psi_{1}(\sigma)\phi_{1}^{\ast}(\sigma')-\psi_{1}^{\ast}(\sigma)\phi_{1}(\sigma'),\\
Y(k)&=-\frac{1}{2}{\rm sech}^2\sigma\left(i\gamma\tan\sigma'+i\gamma\sigma'{\rm sech}^{2}\sigma'+1\right)\notag\\
&\hspace{1 cm}+\frac{1}{2}{\rm sech}^2\sigma'\left(i\gamma\tan\sigma+i\gamma\sigma{\rm sech}^{2}\sigma+1\right)\notag\\
&=-\psi_{1}(\sigma)\phi_{1}(\sigma')+\phi_{1}(\sigma)\psi_{1}(\sigma'),
\end{align}
\end{subequations}
which means ${\cal G}(\sigma)=\delta(\sigma-\sigma')$ and ${\cal R}(\sigma)=0$. Therefore, we obtain
\begin{align}\label{Completeness1}
& |\Phi_{1}(\sigma)\rangle\langle\Psi_{1}(\sigma')|+\int_{-\infty}^{+\infty}dk
|\Phi_k(\sigma)\rangle\langle\Psi_k(\sigma')|
\notag\\
&=I\,\delta(\sigma-\sigma'),
\end{align}
which is (\ref{completeness}) given in the main text for $n=1$.

\section{Atomic spin squeezing}\label{app4}

The Kerr nonlinearity can not only result in the quantum squeezing of the SLDS, but also cause atomic spin squeezing in the system. To show this, we consider the atomic spin operators~\cite{Ma2011,Andersen2016,Schnabel2017}
$\hat{s}_{x}=\frac{1}{2}(\hat{\sigma}_{12}+\hat{\sigma}_{21}),
 \hat{s}_{y}=\frac{1}{2i}(\hat{\sigma}_{12}-\hat{\sigma}_{21}),
 \hat{s}_{z}=\frac{1}{2}(\hat{\sigma}_{11}-\hat{\sigma}_{22}),$
which satisfy the commutation relation
$\left[\hat{s}_{l},\hat{s}_{m}\right]=i\epsilon_{lmn}\hat{s}_{j}$.
We introduce the quadrature spin operator to calculate the spin squeezing
\begin{align}\label{spin2}
\hat{s}_{\theta}&=\frac{1}{2}\left[\hat{\sigma}_{12}e^{-i\theta}+\hat{\sigma}_{21}
e^{i\theta}\right],\notag\\
&=\cos\theta\,\hat{s}_{x}+\sin\theta\,\hat{s}_{y},
\end{align}
and define the minimum spin squeezing degree
\begin{align}\label{spde}
\xi^{2}={\rm min}_{\theta}\left(\frac{\langle\hat{s}_{\theta}^{2}\rangle-\langle\hat{s}_{\theta}
\rangle^{2}}{\langle\hat{s}_{z}\rangle/2}\right).
\end{align}
From the result given by Eq.~(\ref{SE}) and the relation between $\hat{S}_{\alpha\beta}$ and $\hat{\sigma}_{\alpha\beta}$, we can calculate the  minimum spin squeezing degree $\xi^{2}$. Shown in Fig.\ref{Fig6} is
the result of $\xi^{2}$ as a function of propagation distance $s$.
\begin{figure}
\centering
\includegraphics[width=1\columnwidth]{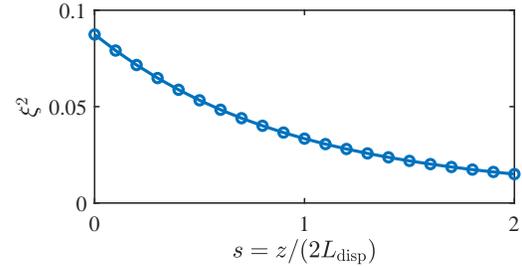}
\caption{~Minimum atomic spin squeezing degree $\xi^{2}$ as a function of propagation distance $s$. }
\label{Fig6}
\end{figure}
We see that the system supports indeed atomic spin squeezing, which is also contributed from the Kerr nonlinearity.



\begin{references}
\bibitem{note1}
Finding a complete and orthommormal set of eigenmodes (consisting of zero modes and continuous modes) for the eigenvalue problem of the soliton perturbation is important for implementing perturbation calculations,
because such eigenmode set provides a complete basis that can express (expand) any perturbation to the soliton in the system.
\bibitem{note2}
If taking $R_0=100~\mu{\rm m}$ (the transverse radius of the probe field), one has the single-photon half Rabi frequency $g_{p}=1.14\times10^{6} ~{\rm s}^{-1}$. By the requirement $L_{\rm disp}=L_{\rm nln}\approx 0.95$\,cm (i.e. $g\approx 1$) for forming soliton, we obtain $n_{0}=9.89\times10^{3}\gg 1$ using the system parameters indicated in the text.
\bibitem{note3}
Here we consider the system with a large size (thermodynamic limit), and there is no external potential. Thus, the Goldstone modes, if they exist, are phonon-like continuous modes.
\end{references}
\end{document}